\documentclass{article}
\usepackage{newcent}
\usepackage{bifchaos}
\usepackage{multicol}
\usepackage{vmargin}
\usepackage{latexsym,float,flafter,balanced,subfigure,longtable}
\usepackage[dvips]{graphicx}

\setpapersize{A4}
\setmarginsrb{1.3cm}{2cm}{2cm}{2.5cm}{0pt}{0mm}{0pt}{0mm}

\newcommand{\BE}{\begin{equation}}
\newcommand{\ee}{\end{equation}}
\newcommand{\BC}{\begin{center}}
\newcommand{\EC}{\end{center}}
\newcommand{\BI}{\begin{itemize}}
\newcommand{\EI}{\end{itemize}}
\newcommand{\BA}{\begin{eqnarray}}
\newcommand{\EA}{\end{eqnarray}}

\begin{document} 

\title{
\bf Boundary Effects in The Complex Ginzburg-Landau Equation
}

\author{
VICTOR M. EGUILUZ\thanks{Corresponding author. E-mail: {\tt
victor@hp1.uib.es}, WWW {\tt http://www.imedea.uib.es/$\sim$victor}} \ ,  
EMILIO HERNANDEZ-GARCIA\thanks{E-mail:
emilio@imedea.uib.es, WWW {\tt http://www.imedea.uib.es/$\sim$emilio}} 
\  and ORESTE PIRO\thanks{E-mail: piro@hp1.uib.es, WWW 
{\tt http://www.imedea.uib.es/$\sim$piro}} \\ 
{\it Instituto Mediterr\'aneo de Estudios Avanzados 
IMEDEA}\thanks{URL: {\tt http://www.imedea.uib.es/Nonlinear}} \
{\it (CSIC-UIB)} \\
{\it E-07071 Palma de Mallorca (Spain)} }

\date{\today}

\maketitle

\vspace{-1.5cm}

\begin{abstract}  
The effect of a finite geometry on the two-dimensional complex 
Ginzburg-Landau equation is
addressed. Boundary effects induce the formation of novel 
states. For example target like-solutions appear as robust solutions 
under Dirichlet
boundary conditions. Synchronization of plane
waves emitted by 
boundaries, entrainment by corner emission, and anchoring of defects by 
shock lines are also reported. 
\end{abstract} 

\vspace{.2cm}

\begin{twocolumns}
\section{Introduction} 
\label{sintro}

The complex Ginzburg-Landau equation (CGL) 
is the generic model describing the slow phase and amplitude 
modulations of a spatially distributed assembly of coupled oscillators 
near its Hopf bifurcation \cite{vSaarloos}. 
It contains much of
the typical behavior observed in spatially-extended nonlinear systems
whenever oscillations and waves are present. After proper scaling it 
can be written as:
\BE
\partial_t A =A  - (1+i\beta)|A|^2 A + (1+i\alpha) \nabla^2 A
\label{cgl}
\ee
where $A$ is a complex field describing the modulations of the 
oscillator field, 
and $\alpha$ and $\beta$ are two real 
control parameters. The first two terms in the r.h.s. of Eq.~(\ref{cgl})
describe the local dynamics of the oscillators:  
the first one is a linear instability mechanism leading to oscillations, 
and the second produces nonlinear amplitude 
saturation and frequency renormalization. 
The last term is the spatial coupling which
accounts both for  diffusion and dispersion of the oscillatory motion.

The power of our analytical tools to study non-linear partial
differential equations in general, and the CGL equation in
particular, is very limited. Roughly speaking, only relatively simple
solutions satisfying simple boundary conditions,  usually in infinite
domains, are amenable to analysis. Examples of these 
are plane and spiral waves. Nevertheless, sustained
spatiotemporally disordered regimes have been 
found and thoroughly investigated numerically. Detailed phase diagrams displaying
the transitions  between different regimes have been charted for
the cases of one and two spatial dimensions
\cite{Shraiman,Chate94,Chate96}. However, we want to stress 
that most of these numerical studies have been performed only under 
periodic boundary conditions, with the underlying
idea that in the  limit of very large systems the boundary conditions
would not influence the overall dynamics. 
As a consequence of this belief, and despite its importance for the 
description of  
real systems, a systematic study of 
less trivial boundary conditions has been largely
postponed. This is the case not only for the CGL equation but also 
for other nonlinear extended dynamical 
systems, and only few aspects of this problem have been collaterally addressed so far 
\cite{Cross80,Cross83,Sirovich}. The purpose of this paper is to report
on the  initial steps of a program aiming towards such a systematic 
study. We will focus here on the behavior of the two-dimensional CGL
equation on domains of different shapes and with different types of boundary
conditions (Dirichlet or Neumann for example). 

For the purpose of comparison 
we first summarize the behavior observed numerically on two-dimensional 
rectangular domains under the commonly used periodic boundary conditions.  
Let us remind that in the so called 
{\it Benjamin-Feir} (BF) {\it stable} region of the parameter space defined
by $1+\alpha \beta>0$, there is always a plane wave  solution
of arbitrarily large wavelength that is linearly stable.
In particular, for parameters in that region, and initializing the system
with a homogenous condition  (a wave of wavenumber $k=0$)
it will remain oscillating homogeneously. If we now vary the parameters
slowly towards crossing the BF line, all the plane waves loss
stability and small perturbations bring the system to a spatiotemporally
disordered cellular state (the so-called {\sl phase turbulence}). It is 
known that the behavior 
close to the BF line can be approximated by the Kuramoto-Sivashinsky equation. 

Further change of the parameters to go deeper
inside the BF unstable region eventually leads to generation of defects,  
i.e., points where $A=0$, and a kind of
turbulent evolution characterized by the presence of these defects sets in.
This is the so called {\sl defect} or {\sl amplitude turbulence}. If
we now trace back to the initial parameter values from the state
dominated by defects, the system does not recover the initial uniformly
oscillatory state.
Spontaneous generation of defects ceases at parameter values still inside the
BF unstable region. At these parameter values, the system usually
reaches a state consisting of a spiral
wave whose core is a defect. This spiral 
occupies most of the domain and it is limited by the shock-lines where the arms of the 
spiral meet themselves. Defects without spiral arms appear at the 
crossings of such shock-lines.
In this regime, the amplitude of the field is time independent and its phase evolves 
quite regularly in time. In general, the configurations that share these two  
properties are called {\it frozen states}. These states persist while we vary 
the parameters all the way back to the
BF stable region. Starting at values corresponding to a defect-dominated evolution,
and  suddenly setting the parameters to values in the stable BF regime, the
stationary solution will be
also a frozen state but in this case several domains, each one
containing a spiral wave, may form. The size of these domains vary
with the initial conditions, but the typical scale is
controlled by the parameters. Shock lines where the arms of different
spirals collide now proliferate and non-spiral defects are usually present 
at the crossings between them.

\section{Boundary effects}

Let us consider first parameter values such that with periodic boundary 
conditions  the long-time asymptotic 
states are {\sl frozen} and look at how  
the behavior is modified by  changing the
boundary conditions. We apply null Dirichlet ($A=0$), and
Neumann (vanishing of the normal derivative of $A$) boundary
conditions. For the former, we consider three different boundary shapes:
square, circle, and stadium-shaped domains. Comparison between square and 
circle will allow us to investigate the influence of corners. On the other 
hand, our interest in the
stadium arose from considerations of ray chaos, but it 
will be presented here as a
combination of circle and square geometries. 

In the Dirichlet case, 
the zero amplitude boundaries 
facilitates the formation of defects near the walls. Starting from 
random initial conditions, defects are actively 
created in the early
stages of the evolution. After some time however 
all the points on the boundaries synchronize and
oscillate in phase so that plane waves are emitted. Defect formation
ceases, and the waves emitted by the walls push
the remaining defects towards the central region of the domain. There 
the defects annihilate in pairs of opposite charge and as a
result of this process a bound state is formed 
by the surviving set of equal-charge defects.  The orientation of the
waves emitted by the boundaries also changes during
the evolution. The synchronized emission of the early stages
proceeds, obviously, perpendicular to the boundary  but later 
the wavevector tilts to some emission angle
of approximately $45$ degrees. This angle depends on both the parameter
values and the geometry of the boundaries. The fact that this angle is not
exactly $45$ degrees
is made evident by a mismatch of the waves coming from orthogonal 
walls. Finally the system reaches a frozen state of the type displayed in 
Fig. 1. The defects are confined 
in the center of the domain forming a rigid static chain. The constant-phase 
lines travel from the boundaries towards the center of the domain. Shock 
lines appear where waves from different sides of the contour collide. The 
strongest shocks   
are attached perpendicularly to the walls. If for a particular initial 
condition  
all defects annihilate the asymptotic state is a defect-free 
{\it target} solution. This kind of solutions is not seen  
seen in simulations with periodic boundary conditions.

\begin{figure}[H] 
\begin{center}
\resizebox{75mm}{!}{\includegraphics{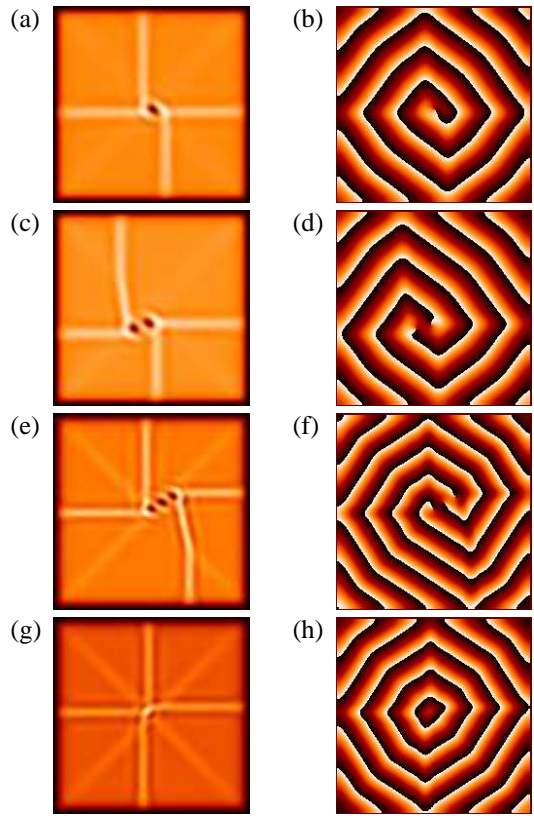}}
\end{center}
\caption{Frozen structures under null Dirichlet boundary 
conditions in a square of 
size $100\times100$. Parameter values are $\alpha=2$, $\beta=-0.2$ (a-d), 
and $\alpha=2$, $\beta=-0.6$ (e-h).  
Snapshots of the modulus $|A|$ of the field 
are shown in the left column and snapshots of the phase in the right column. Color
scale runs from black (minimum) to white (maximum).}
\label{ffrozen_s}
\end{figure}

It is known \cite{Hagan} that the phase velocity of the usual spiral waves 
in infinite systems could point
either inwards or outwards the defect core depending on the parameter values. 
In our simulations in the square geometry 
with Dirichlet conditions, however, 
the direction of the phase velocity is always from the
boundary to the core. We can understand this better by applying null 
Dirichlet conditions to only one of the walls. The synchronized 
emission that we observe is a straightforward generalization
to  two-dimensions of the one-dimensional
Nozaki-Bekki emitting hole solution \cite{Nozaki}. 
We have verified \cite{Eguiluz}, for instance, that 
the direction of the emitted waves (inwards or outwards)  
can be changed with parameters as predicted by
the analytic computations \cite{Hagan}. However, when several of the walls 
are lines of zeros (the four sides of the square, for example) 
the direction of the phase velocity becomes determined by the angle 
between these lines. In other words, corners effectively entrain the 
whole system. 

\begin{figure}[H]
\begin{center}
\resizebox{75mm}{!}{\includegraphics{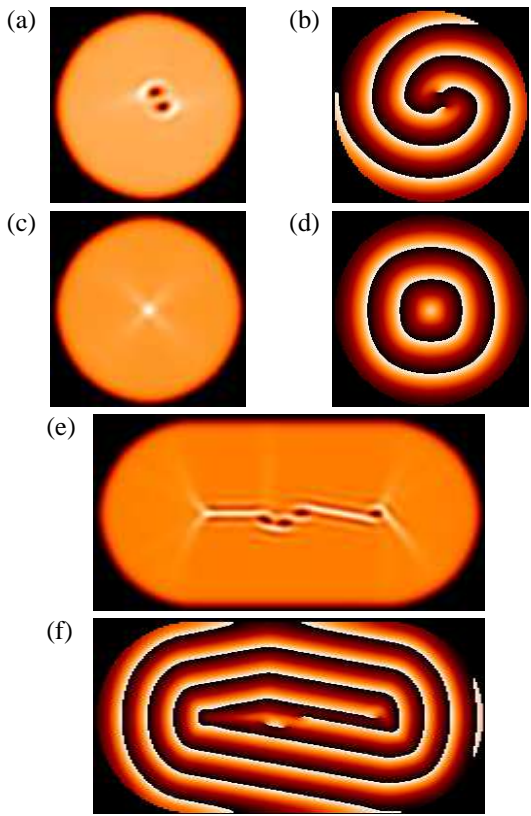}}
\end{center}
\caption{Frozen structures under null Dirichlet boundary conditions 
in a circle (a-d) 
of diameter $100$ for parameter values 
$\alpha=2$, $\beta=-0.2$, and  
in a stadium (e-f) of size $200\times100$, for parameter values
$\alpha=2$, $\beta=-0.6$.  
Snapshots of the modulus $|A|$ are shown in the left column and (e) whereas 
the phase is shown 
in the right column and (f). Color scale as in Fig.~1.}
\label{ffrozen_c}
\end{figure}

In a circular domain  (Fig.~\ref{ffrozen_c}), the frozen structures are
either targets (no defects) or a single central defect. Groups of defects of
the same charge can also form bound states, but instead of freezing they  
rotate together. This contrasts with the behavior
of the square domains and is  
correlated with the absence of shock
lines linking the boundaries to the center in the case of the
circular domains. These links are probably responsible for 
providing rigidity to the stationary
configuration in the square case. Tiny shock lines associated
to small departures from circularity in the lines of constant phase can
be observed also in the circle but 
these lines end in
the bulk of the region before reaching the boundaries.
On the other hand, the constant-phase lines reach the boundaries
nearly
tangentially in contrast to what we observe in the square.
In addition, we observe that for circular domains the phase velocity
direction can be changed controlling the parameters. This is
probably a consequence of the absence of the corners
that synchronize the emission from the boundaries in
the square case. 

\begin{figure}[H]
\begin{center} 
\resizebox{75mm}{!}{\includegraphics{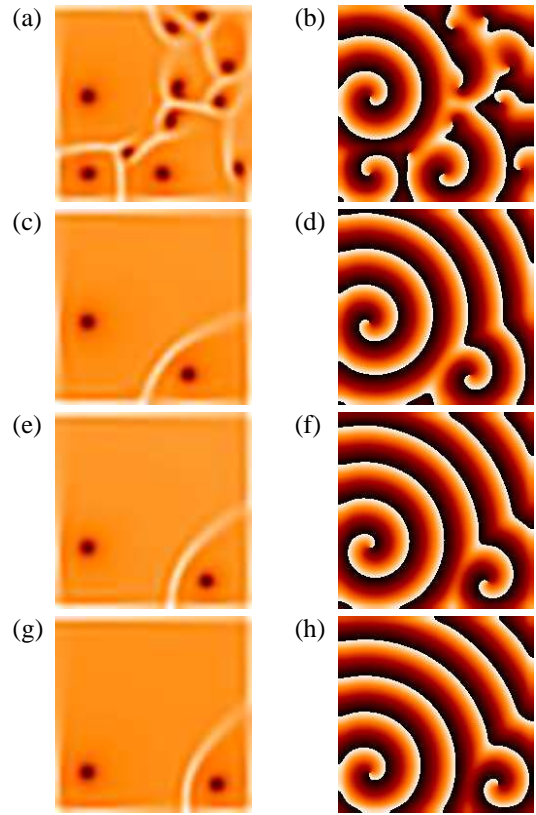}}
\end{center} 
\caption{Snapshots
of the field $|A|$ (left column) and phase (right
column) in color scale as in Fig.~1 at times 
$t=2.5 \times 10^4$ (a-b), $t=5.\times 10^4$ (c-d), 
$t=7.5 \times 10^4$ (e-f), and  $t=10. \times 10^4$ (g-h) under 
Neumann boundary conditions in a square domain  
of size $100\times100$. Parameter values are $\alpha=2$, $\beta=-0.2$. } 
\label{fneumann}
\end{figure}

The stadium shape (Fig.~\ref{ffrozen_c}) mixes features of the
two geometries  previously studied: it has both straight and 
circular borders. In this case, the curves of constant phase
arrange themselves to combine the two behaviors
described above. On one hand the lines meet
the straight portions of the border of the stadium with some
characteristic angle, as it happens in
square domains. However, these lines bend to become nearly tangent
to the  semicircles in the places where they meet with these
portions of the boundaries.
A typical frozen solution displays a shock
line connecting the centers of the circular portions of the domain. 
This shock  line
usually contains defects. It is also possible to find
defect-free target solutions as in the case of the circle, and the 
behavior of the phase velocity is also similar in the sense that
its direction can be changed by modifying the parameters.

The behavior under Neumann boundary conditions 
is rather similar to the case  of periodic boundary 
conditions. However,  the Neumann conditions induce
several subtle features to the dynamics. For example,
shock lines are now forced to reach orthogonally 
the boundaries. In addition,
defects can be irreversibly absorbed  by the boundaries, 
a process that is obviously impossible with periodic
boundary conditions. During the evolution a spiral defect
behaves as if it 
were interacting with a mirror image of itself
with opposite charge located outside the domain \cite{Aranson93}.
This reflects in few characteristic
phenomena. On one hand an isolated
defect tends to move parallel to a nearby Neumann wall.
On the other hand, mutual annihilation of a defect and its image
is also possible accounting for the absorption of this defect by the
boundary.
Finally, when a defect closely approaches a corner, its evolution
gains in complexity possibly as a result of the mutual interaction
with two different images. 
Fig.~\ref{fneumann} displays a typical evolution of the pattern.
Initially starting at random, a number of  dynamically active spiral
defects is created. These move around eventually annihilating
mutually or sometimes being absorbed by the walls while the
dynamics progressively slows down. Normally one large spiral
wave grows until it fills
the whole domain at the expense of the smaller ones that
are pushed out of the boundaries.

Finally, we have studied the changes induced by
the boundaries for parameter
values such that active spatiotemporal chaos (i.e., non-frozen states) 
is found for periodic boundary conditions.
Far from the boundaries 
spatiotemporally chaotic solutions behave
similarly to those satisfying periodic boundary conditions.
However, a boundary layer with different behavior shows up near the 
borders. In Fig.~\ref{fdyn} we can see plane waves
emitted by the boundaries and rapidly fading inside
the domain where spatiotemporal chaos evolves. In 
small domains the boundaries could synchronize the whole 
system. However, as the system size increases,
full synchronization ceases. 

For other parameter values, Dirichlet boundary conditions lead 
eventually to a dynamics characterized by the
coexistence of regions dominated by defect turbulence
and regions dominated by
plane waves (constant $|A|$) whose shape and position
normally evolve in
time. We have found this behavior in all the domain shapes
studied except for
circular case.  

For these parameter values, Neumann boundary conditions
do not produce a dynamics sensibly different than
the one induced by periodic boundary conditions. 
The only noticeable difference is that in the Neumann case 
the shock lines 
are forced, as pointed out before, to meet orthogonally
the boundaries.

\section{Conclusions} 

In this paper, we have presented important features of the dynamics
of the CGL equation which depend strongly on the type of boundary conditions imposed,
as well as on the geometrical shape of the boundaries.

\begin{figure}[H] 
\begin{center}
\resizebox{75mm}{!}{\includegraphics{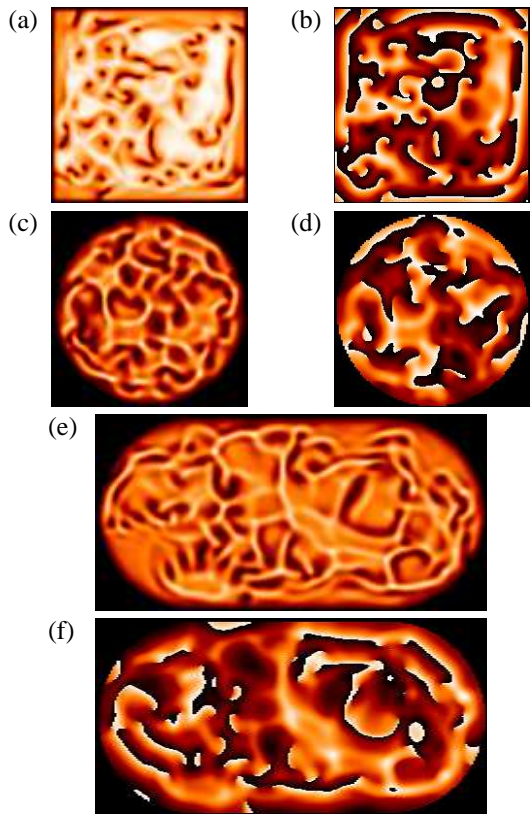}}
\end{center}
\caption{Dynamical solutions under Dirichlet boundary 
conditions. 
Snapshots of the field $|A|$ are shown in the left column and (e) 
whereas the phase is shown in the right column and (f). 
(a-b): square, parameter values $\alpha=2$, $\beta=-0.75$; 
(c-d): circle, parameter values $\alpha=2$, $\beta=-1.$; 
(e-f): stadium, parameter values $\alpha=0$, $\beta=1.8$. System 
sizes and color scale as in Figs. 1 and 2. }
\label{fdyn}
\end{figure}

Dirichlet boundary conditions play a double r\^ole. On one hand, 
the walls naturally behave as sources (or sinks) of defects.
On the
other hand, a wall with null Dirichlet conditions shows a tendency to emit 
plane waves. The interplay
between these two properties of the boundaries gives rise to 
interesting behavior. 

In the case of frozen states, the character of the walls as wave 
emitters dominates. 
Some geometrical features of the boundaries have a strong
influence on the details of the phase synchronization.
Corners, for instance, tend to act as pacemakers. 
In circular domains, on the other hand, the emission is
definitively dominated by the internal spirals. 
Correspondingly, the internal structure of
the frozen states is also influenced by the shape of the
boundaries. In a square, defects form a chain which is 
anchored to the boundaries by a set of
shock lines; in a circle, on the contrary, the asymptotic
state is usually a bound state disconnected from the boundaries. 

Neumann boundary conditions seem to have a much weaker influence
on the overall dynamical behavior of the CGL equation. However 
some differences are evident: One is the orientation of the shock lines, 
perpendicular to the boundaries. The other is that 
defects can be ejected through the boundaries, thus favoring 
states dominated by a single spiral in situations where under 
periodic boundary conditions a {\sl glassy} state with several 
spiral domains would be formed.

Since the CGL equation appears naturally in a variety 
of contexts,  
we believe that the phenomena found
in our preliminary explorations are likely to be relevant
in many theoretical and experimental situations. Some of the
phenomena reported here have intrinsic interest 
and deserve further analysis.

\section{Acknowledgments}

Financial support from Direcci\'on General de Investigaci\'on 
Cient\'{\i}fica y T\'ecnica (DGICYT, Spain), Projects PB94-1167 and 
PB94-1172.

\bibliographystyle{bifchaos}

\end{twocolumns}

\end{document}